\documentclass[3p,times]{elsarticle}

\usepackage{ecrc}

\volume{00}

\firstpage{1}

\journalname{Journal of Systems and Software}

\runauth{}

\jid{procs}

\jnltitlelogo{Procedia Computer Science}

\CopyrightLine{2011}{Published by Elsevier Ltd.}

\usepackage{amsmath}
\usepackage{amssymb}
\usepackage{amsthm}
\usepackage{multirow}
\usepackage{array}
\usepackage{lineno}
\usepackage{amssymb}

\usepackage{longtable}

\newdefinition{rmk}{Remark}
\newproof{pf}{Proof}
\newproof{pot}{Proof of Theorem \ref{thm2}}

\usepackage[figuresright]{rotating}

\begin{document}

\begin{frontmatter}



\dochead{}

\title{BotHawk: An Approach for Bots Detection in Open Source Software Projects}


\author[aff1]{Fenglin Bi}
\ead{52215903006@stu.ecnu.edu.cn}

\author[aff1]{Zhiwei Zhu}
\ead{634262410@qq.com}

\author[aff1]{Wei Wang}
\ead{wwang@dase.ecnu.edu.cn}

\author[aff1]{Xiaoya Xia}
\ead{wwang@dase.ecnu.edu.cn}

\author[aff2]{Hassan Ali Khan}
\ead{hassankhanzae@gmail.com}

\author[aff1]{Peng Pu\corref{cor1}}
\ead{ppu@cc.ecnu.edu.cn}

\cortext[cor1]{Corresponding author}

\address[aff1]{School of Data Science and Engineering, East China Normal University. Shanghai, China}
\address[aff2]{Institute of Software Engineering, East China Normal University. Shanghai, China}

\begin{abstract}
Social coding platforms have revolutionized collaboration in software development, leading to using software bots for streamlining operations. However, The presence of open-source software (OSS) bots gives rise to problems including impersonation, spamming, bias, and security risks. Identifying bot accounts and behavior is a challenging task in the OSS project. This research aims to investigate bots' behavior in open-source software projects and identify bot accounts with maximum possible accuracy. Our team gathered a dataset of 19,779 accounts that meet standardized criteria to enable future research on bots in open-source projects. We follow a rigorous workflow to ensure that the data we collect is accurate, generalizable, scalable, and up-to-date. We've identified four types of bot accounts in open-source software projects by analyzing their behavior across 17 features in 5 dimensions. Our team created BotHawk, a highly effective model for detecting bots in open-source software projects. It outperforms other models, achieving an AUC of 0.947 and an F1-score of 0.89. BotHawk can detect a wider variety of bots, including CI/CD and scanning bots. Furthermore, we find that the number of followers, number of repositories, and tags contain the most relevant features to identify the account type.
\end{abstract}

\begin{keyword}
Dataset \sep Bots Detection \sep Classification \sep Open-source Software Bots 


\end{keyword}

\end{frontmatter}


\section{Introduction}
\label{S:1}

The growth of the software development industry has been unprecedented, and the surge in paid development has necessitated extensive evaluation of software functions. Consequently, collaboration, as a social phenomenon, has become increasingly critical in the software development lifecycle. Prevalent social coding platforms, such as GitHub, Bit-Bucket, and GitLab, facilitate collaboration by providing a shared workspace for developers to collaborate on projects \cite{dabbish2012social}. As a result, developers can more effectively share knowledge, identify and fix bugs. Collaboration among developers provides an opportunity to learn from each other, broaden their understanding of diverse programming languages and methodologies, and ensure that the software satisfies high quality standards while being delivered promptly. The pull-based development process, utilized in distributed development platforms like GitHub, is the primary method for integrating code from numerous developers \cite{gousios2014exploratory}. While this mode has significantly impacted the development of open-source software, it has also massively increased the workload for repository maintainers, requiring them to manage communication with contributors, review source code, handle contributor license agreement issues, explain project guidelines, run tests and build code, and merge pull requests. The described process had a considerable impact on the evolution of open-source software. However, it also resulted in a substantial increase in workload for repository maintainers \cite{gousios2016work}. 

In an effort to alleviate the burden of repetitive tasks, open-source software (OSS) projects have recently utilized various software bots to streamline their operations. OSS project management relies on bots to oversee collaborations and prevent coding errors on hosting platforms by performing tasks like ensuring license agreement signing (e.g., Googlebot) and running automated tests (e.g., pdf.js test). As an interface between humans and services, bots play a crucial role in social coding platforms \cite{storey2016disrupting}.  

While OSS bots can help with repository maintenance, they also display social bot characteristics that can cause problems when used for software development activities. These include: 1) Impersonation: The accuracy of open-source empirical research is hindered by the presence of bot behavioral data. \cite{xia2022exploring}. 2) Spamming: Bots can create a large volume of low-quality or irrelevant content, such as comments or pull requests, which can overwhelm the repository and distract developers from more crucial tasks \cite{shao2018spread}. 3) Bias: Bots may introduce bias into the development process by favoring specific types of contributions or contributors over others \cite{gao2022understanding}. 4) Security risks: Malicious bots can launch attacks on the repository or steal confidential information, posing a significant security risk \cite{pearce2022asleep}. Therefore, many open-source researchers need to identify OSS bot accounts and behavior. For example, Maldeniya et al. investigated the composition and operation of virtual, loosely-knit teams in the collaborative crowdsourcing model of OSS development. To improve research accuracy, they excluded the activities of automated accounts that are discernibly different from traditional organizations \cite{maldeniya2020herding}. By studying the behavior and impact of bots in software development, researchers can gain insights into how to mitigate these challenges and optimize the use of bots. Researchers have collected various datasets to perform clear classification tasks, and machine learning models such as random forests and neural networks have achieved near-perfect performance. OSS bot detection techniques are in general not made public, researchers, journalists, and the public at-large rely on researcher-developed tools to separate bots from genuine human users and understand the impact of bots on social phenomena \cite{hays2023simplistic}. 

Identifying bots is a challenging task due to their actions' similarity to those of a human. This is compounded by the ability to trigger a bot's activity through a platform's API or directly on the platform's website \cite{chu2012detecting}. In addition, Identifying OSS bots is more challenging than identifying bots on social media due to the complexity of their functions and their participation in various open-source collaboration events \cite{schueller2022evolving}. The blending of automated and human behavior further adds to the difficulty of detection \cite{pozzana2020measuring}. We find that there are a limited number of attempts to identify and classify OSS bots.

Evidence suggests that bot detection tools remain imperfect as they are either unreliable over time or rely on dubious features. Furthermore, researchers do not consider the problem solved, as earlier research has uncovered the possibility of bot classifiers' failure, and there are concerns that more sophisticated bots may remain undetected.  \cite{efthimion2018supervised}.   

Evaluating OSS bot detection datasets and models is a challenging task. The "ground truth" is either unknown or inaccessible to the public, and our understanding of OSS bots is limited to the datasets themselves. Moreover, different datasets may lack sufficient evaluations of OSS bot models. Nonetheless, this challenge does not render the evaluation process impossible. We can gain a more profound comprehension of the information by examining the datasets. 

We propose \textbf{BotHawk}, an extensive approach to identifying bots and creating a ground truth dataset that conforms to the real world. Our dataset is more representative and includes a broader range of OSS bot types. To improve our identification approach, we categorized OSS bots according to their behaviors and extracted 17 properties. We evaluated BotHawk compared with state-of-the-art open-source bot detection methods and demonstrated its superiority in terms of identification metrics and overall performance. And we analyzed the factors contributing to the superior performance of the BotHawk model in detecting open-source bots and identified the most important three features. The source code and dataset for this project are open source and available on GitHub at https://github.com/bifenglin/BotHawk.

The rest of this paper is organized as follows: Section 2 presents motivation and related work. Section 3 explains how we created the ground-truth dataset. Section 4 outlines the features we selected for the classification model. Section 5 details the workflow for selecting and evaluating an appropriate classification model. Section 6 discusses the results, and Section 7 concludes the paper.

\section{Motivation and related work}
\label{S:2}

The increasing prevalence of bots in open-source software projects has prompted extensive research on bot detection techniques. While various approaches have been proposed, limitations regarding accuracy, scalability, and adaptability still need to be made. In this section, we discuss the motivation behind our work, highlighting the gaps in existing research and emphasizing the need for a more robust bot detection solution. Furthermore, we provide an overview of related work, comparing and contrasting different methods and techniques that have been employed in previous studies.  

\subsection{Motivation}

Numerous researchers have extensively studied bot detection. The underlying motivations include (1) Data Cleaning: Some bot behavioral data can be perplexing in exploring and evaluating accuracy due to invalid, duplicate, or redundant parts. (2) Expanded Bot Research: Discover bot-related data and conduct additional research on OSS bots. (3) Platform Maintenance: As a global open-source community, upholding a healthy and orderly development environment is imperative. A plethora of irrelevant or malicious bot accounts can have an adverse impact on the platform order, leading to a breakdown in the developers' trust.  

Bot accounts' behavioral data can be inclined and impact data-driven research  \cite{golzadeh2021ground}. Dey et al. \cite{dey2019patterns} demonstrated how bots generated numerous issues and pull requests in the under investigation code repository, which could disrupt research results and conclusions. Developers seek techniques to detect bots on GitHub and exploring prevalent Q\&A websites \cite{stackoverflow2019cga}. Human-centered software maintenance studies must accurately differentiate between human and bot activity since misrepresenting the nature of the activity could potentially compromise the internal validity of a study. Consequently, many software engineering studies have excluded bots and bot activity \cite{peng2018exploring}.  

In the open-source community, several researchers have conducted extensive studies on bots used in OSS. These studies have focused on various aspects, including the identification and classification of bot accounts, analyzing behavior patterns associated with bot accounts, and examining the impact of these accounts on the community and developers. Through these studies, a better understanding of the role and influence of bot accounts in the GitHub community can be obtained, providing guidance and support for better management and use of bot accounts. At the same time, it can also provide references and inspiration for bot research in other fields. 

As one of the world's largest source code hosting platforms, GitHub is also an important channel for many open-source projects to operate. By identifying bot accounts, these activities can be better managed and controlled, and the effectiveness of these activities can also be better monitored and evaluated. If there are a large number of invalid or malicious bot accounts, it will have a negative impact on the platform's order and even affect the trust of developers \cite{souza2021analysis}.

\subsection{Related works}
Research in this field primarily concentrates on three areas: (1) the classification of open-source bots, (2) datasets for bot detection and feature extraction in open-source software, and (3) algorithms for bot detection.  

The researchers have classified software bots from various perspectives. Lebeuf proposed a faceted taxonomy for software bots \cite{lebeuf2019defining}. "software bots" is inclusive and not solely focused on software engineering activities, components, and roles. Lebeuf's taxonomy is extensive and divided into three dimensions, which cover a total of 22 aspects. These dimensions include the bot's environment, internal properties exhibited by the bot, and the interaction between the bot and its environment. However, their taxonomy is relatively complex for bots in the open-source domain. Wessel et al. \cite{wessel2018power} have conducted significant research to open-source software bots. They acquired 351 popular open-source projects from the GitHub platform. They detected 93 of them (26\%) utilizing bots to execute automated predefined and repetitive tasks, supporting the work of the developers and contributors. These bot operations have been categorized into functions such as "Ensuring License Agreement Signing" and "Reporting Continuous Integration Failures." However, their classification method is less useful for identifying bots using automated tools. Erlenhov et al. \cite{erlenhov2019current} identified the characteristics of DevBots (bots that support software development) by applying an aspect-based taxonomy. They provide examples of this taxonomy, utilizing 11 established, industrial-strength bots. Their taxonomy is limited to bots that support software development and does not extend to the entire domain of open-source software bots.  

The original dataset utilized for open-source software bots was predominantly processed from GitHub API, the GHTorrent dataset, and the World of Code dataset. Golzadeh et al. \cite{golzadeh2021ground} selected 136K software package registries, including PyPI's GitHub repositories. They excluded the accounts with fewer than ten comments, leading to 79,342 GitHub accounts. Finally, the authors manually annotated 5,082 accounts based on their discourse within issues/pull requests. The resulting dataset comprises 5,000 GitHub accounts, with 4,473 human accounts and 527 bot accounts. However, their features are limited, primarily using comment data from issues. According to Zhao et al. \cite{zhao2017impact}, time-series-related features are also significant, but they were not considered in their study. Dey et al. \cite{dey2020detecting} generously shared a bot dataset, which comprises 461 bot accounts and an extensive collection of commits totaling 13,762,430. The dataset includes submission metadata, account names, and email addresses. However, the account login names for GitHub accounts are absent from the dataset and they lack time-series-related features. Several other datasets explain their methods of construction but have not been made public. For example, \cite{abdellatif2022bothunter} combines two publicly available datasets and filters out inactive accounts since 2017. In \cite{wessel2020effects}, The authors created time-series by considering projects that have been active for at least one year before and after adopting bots. The authors identify 4,767 projects manually where at least one code-review bot has been employed. DevBots dataset \cite{erlenhov2019current} was discovered through Internet search and social media advertising, and only 11 bot accounts were found.

In recent years, numerous methods have been proposed to differentiate between human and bot comments on social programming platforms such as GitHub. Cassee et al. \cite{cassee2021human} studied three machine learning classifiers to recognize pull requests and issue comments submitted by artificial agents. Golzadeh et al. \cite{golzadeh2021ground}  presented BoDeGHa, a machine learning-based approach that identifies software bot account on GitHub by analyzing comment-related features like repetitive comment patterns. The authors evaluated their proposed technique on 5000 GitHub accounts.  In their subsequent research \cite{golzadeh2021identifying}, the authors expanded the classification to the comment level to determine the origin of comments as bot-generated or human-generated. They introduced a classification model that relied on natural language processing techniques. The foundation of their study was a well-balanced ground-truth dataset encompassing 19,282 comments derived from pull requests and issue discussions. To convert the comments into vector representations, they employed a hybrid approach that combined the bag of words and TF-IDF approach. Ahmad Abdellatif et al. proposed BotHunter \cite{abdellatif2022bothunter}, a machine learning-based method for identifying bot accounts. This method utilizes 19 pre-selected features, including personal profile information, activity patterns, and comment similarity. The researchers evaluated the effectiveness of these features by conducting experiments with a dataset containing over 5000 GitHub accounts. In a separate study, Dey et al. \cite{dey2020detecting} developed BIMAN, a system designed to detect bots that submit code by analyzing submission metadata, particularly modified files. The performance of the proposed methods was assessed using different evaluation metrics. The F1 score attained by BoDeGHa was 98\%, whereas BotHunter achieved 92.4\%. Moreover, BIMAN yielded an AUC-ROC of 90\%.

In light of the existing research, our work seeks to address the identified limitations by proposing BotHawk, a more robust and adaptable bot detection approach tailored to the unique challenges posed by the open-source software ecosystem.

\section{Ground truth dataSet}
\label{S:3}

To develop and evaluate our BotHawk approach, we rely on a ground truth dataset of labeled bot and genuine user accounts from open-source software projects. In this section, we describe collecting, curating, and labeling the dataset, ensuring that it accurately represents the diverse range of bots and user activities present in the open-source ecosystem. We also discuss the challenges faced during the dataset creation process and the steps taken to mitigate potential biases and inaccuracies.

Real-world datasets are often "dirty" and plagued with various data quality issues.  However, data quality is crucial for ensuring that machine learning systems that rely on data can accurately represent and predict phenomena. An increasingly recognized and growing body of research focuses on understanding and improving data quality \cite{hazelwood2018applied}. Moreover, data cascades can impede research on open-source software bot detection \cite{sambasivan2021everyone}. The text data-based random forest algorithm achieved almost perfect performance on bot recognition dataset \cite{golzadeh2021ground}. How can we reconcile these results, which contradict our intuition that bot detection is a challenging issue? On the one hand, it is plausible that bot detection is simpler than expected, and simple decision rules suffice. On the other hand, the datasets may not capture the true complexity of bot detection. This suggests that simple decision rules may perform well in sample but exhibit significantly lower performance when deployed. We use a more comprehensive dataset to test advanced models to prove the latter hypothesis.  

When creating a comprehensive dataset for open-source software bot research, the following aspects should be considered: (1) data accuracy. The bot detection feature and label data in the dataset must be precise to ensure the validity and dependability of subsequent machine learning models. (2) Generalization ability - The dataset ought to cover all possible bot detection variables and information for machine learning models. (3) Data Scalability. It is important to create a scalable dataset to accommodate new data at any time, hence improving the efficiency and effectiveness of machine learning models. (4) Data Timeliness. Seeing as data originates from real-world bot activities, it is necessary to keep updating the data in real time to prevent it from becoming invalid. In addition to the above, the dataset's distribution must be considered to ensure that it conforms with practical application scenarios' requirements, thereby enabling machine learning models to respond efficiently to practical challenges.  

\begin{figure}[h]
\centering\includegraphics[width=1\linewidth]{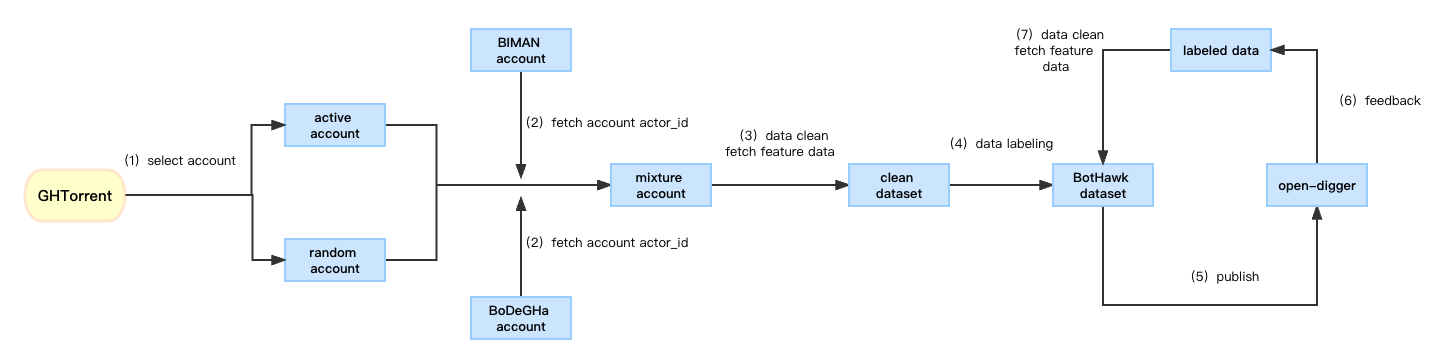}
\caption{Workflow of BotHawk dataset process}
\label{fig:Workflow of BotHawk dataset process}
\end{figure}

\begin{figure}[htbp]
\centering
\begin{minipage}[t]{0.48\linewidth}
\centering
\includegraphics[width=\linewidth]{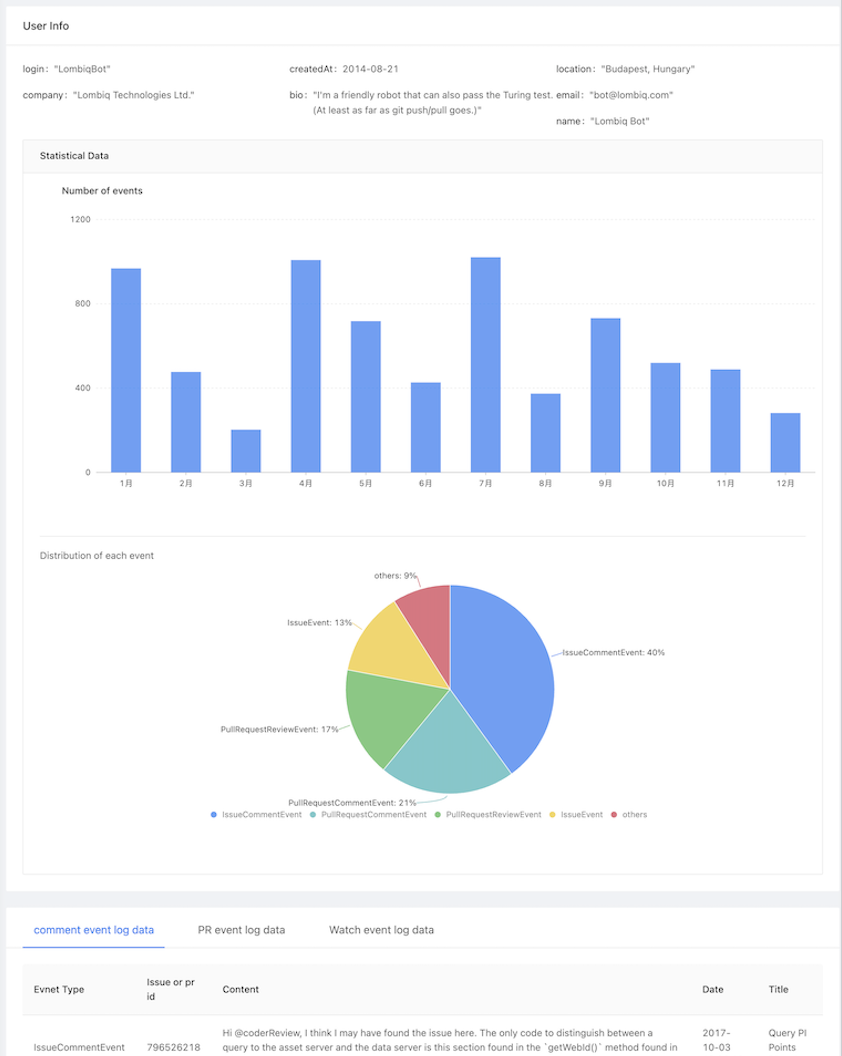}
\caption{Anonymised screenshot of the label application}
\label{fig:Anonymised screenshot of the label application}
\end{minipage}
\hfill
\begin{minipage}[t]{0.48\linewidth}
\centering
\includegraphics[width=\linewidth]{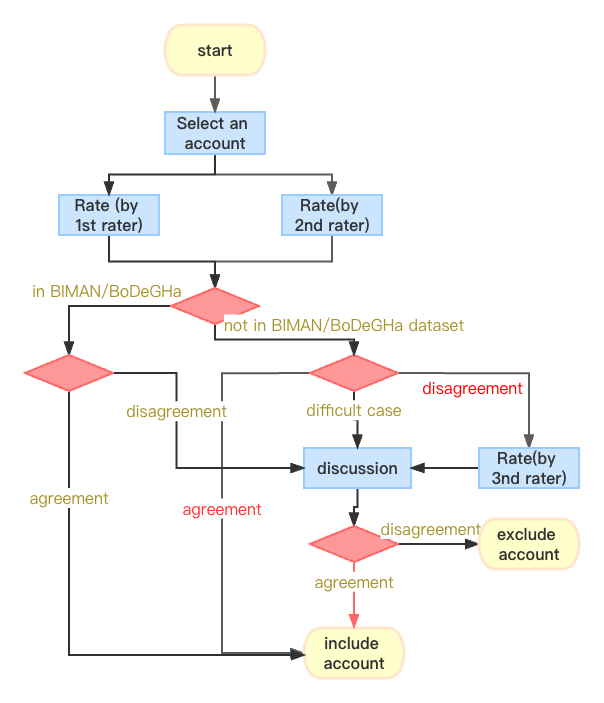}
\caption{Workflow of labeling process}
\label{fig:Workflow of labeling process}
\end{minipage}
\end{figure}

The process of the BotHawk dataset is depicted in Figure~\ref{fig:Workflow of BotHawk dataset process}. (1) To ensure generalization ability and accuracy, we selected the most active repositories from the GHTorrent dataset between March 2021 to March 2022. We identified "active accounts" with activity data that exceeds 100 logs. Then, we randomly selected a portion of accounts from the global accounts as "random accounts." (2) To expand the dataset and ensure the credibility of the comparative experiments with the BIMAN and BoDeGha algorithms, we processed the data from BIMAN and BoDegHa to obtain their account's GitHub IDs, and selected accounts that have been active in the past year (with activity data greater than 10 logs). We combined the data from "active accounts," "random accounts," "BIMAN accounts," and "BoDeGha accounts" into a single account set called "mixture accounts." (3) We cleaned the "mixture accounts" and selected 17 relevant features to ensure the comprehensiveness of the data, which constituted the "clean dataset." We will elaborate on the feature selection process in section 4. (4) We obtained and labeled the Bothawk dataset using features and their activities on GitHub. To ensure label accuracy, we developed a data query and visualization application (Figure~\ref{fig:Anonymised screenshot of the label application}) and defined a set of data labeling processes (Figure~\ref{fig:Workflow of labeling process}). Our approach consolidates the data labeling process and enhances the reliability of the dataset. (5) We uploaded the labeled data to the opendigger project, which is a crowdsourced GitHub repository offering labeled open-source datasets. This step was taken to enhance the timeliness and accuracy of the dataset. Eventually, we plan to share our data processing codes on GitHub and update the dataset monthly to enhance its scalability and timeliness. 

We have developed a comprehensive BotHawk dataset, consisting of 19,779 accounts with 17 distinct features, to facilitate the detection of bots in open-source software projects. The dataset is carefully constructed to include a diverse dataset, as shown in Figure~\ref{fig:Each dataset proportion in Bothawk}. The reason why the data volume of BIMAN is the smallest can be attributed to the fact that we have associated more than 2000 pieces of data with GitHub accounts through email accounts. However, a considerable portion of them has been inactive lately. Consequently, we have not considered accounts with less than 10 activity logs while compiling our dataset.  

The carefully curated ground truth dataset serves as the foundation of our research, enabling us to train and evaluate the BotHawk model in a reliable and representative manner, ultimately improving its generalizability and real-world applicability.

\section{Feature selection}
\label{S:4}

This section presents the feature selection process employed to identify the most relevant features for detecting bots in open-source software projects. Feature selection is a crucial step in building an effective machine learning model, as it helps eliminate redundant or irrelevant features, reduce the dimensionality of the dataset, and improve the model's performance. We begin by discussing the initial set of features and then introduce the feature selection techniques used to refine the feature set.

\subsection{Introduction to Feature Selection}  
In the realm of machine learning, feature selection plays a crucial role in improving the performance and interpretability of models, as well as reducing their complexity. In the context of our BotHawk project, selecting the most informative and discriminative features is vital for accurately detecting bots in open-source software projects. This section will explore identifying relevant features by analyzing bot behavior and drawing upon the BotHawk dataset.

The software development process involves a diverse range of roles for software developers, and similarly, plenty of bots participate in each phase of the process \cite{storey2016disrupting}. How can we comprehensively detect OSS bot accounts as thoroughly as possible? Relying solely on analyzing comments is not sufficient to detect OSS bots. In addition to responding to comments, the lifecycle of these bots extends throughout the entire code management and software development process in the collaboration workflows of the platform.   

The diversity of OSS bots poses a challenge to researchers and developers seeking to compare and evaluate the different types of bots. Research in this area necessitates a multidisciplinary understanding, including knowledge of open-source collaboration patterns, DevOps, and legal fields. Analyzing OSS bots requires substantial data analysis, which can be challenging, particularly in data acquisition and processing. Our research relies on GitHub data, as it is the most widely used code hosting platform. \cite{cosentino2016findings}. With over 94 million developers and 90\% of Fortune 100 companies using the GitHub platform, the number of GitHub behavior logs is expected to reach 100 million rows by 2023 \cite{jeremy2023key}. This vast amount of data requires efficient methods for analysis and processing. To collect our data, we primarily rely on the GH Archive project, a popular GitHub data-sharing repository that many researchers use, and the GitHub API, which is an application programming interface that provides developers access to the data stored on GitHub.

\begin{figure}[htbp]
\centering
\begin{minipage}[t]{0.48\linewidth}
\centering
\includegraphics[width=\linewidth]{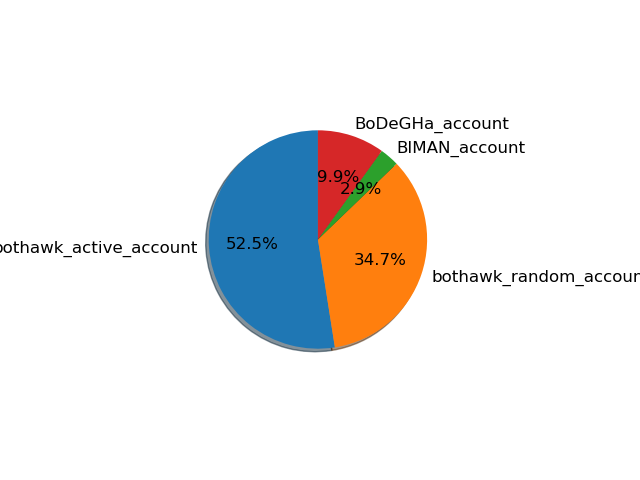}
\caption{Each dataset proportion in Bothawk}
\label{fig:Each dataset proportion in Bothawk}
\end{minipage}
\hfill
\begin{minipage}[t]{0.4\linewidth}
\centering
\includegraphics[width=\linewidth]{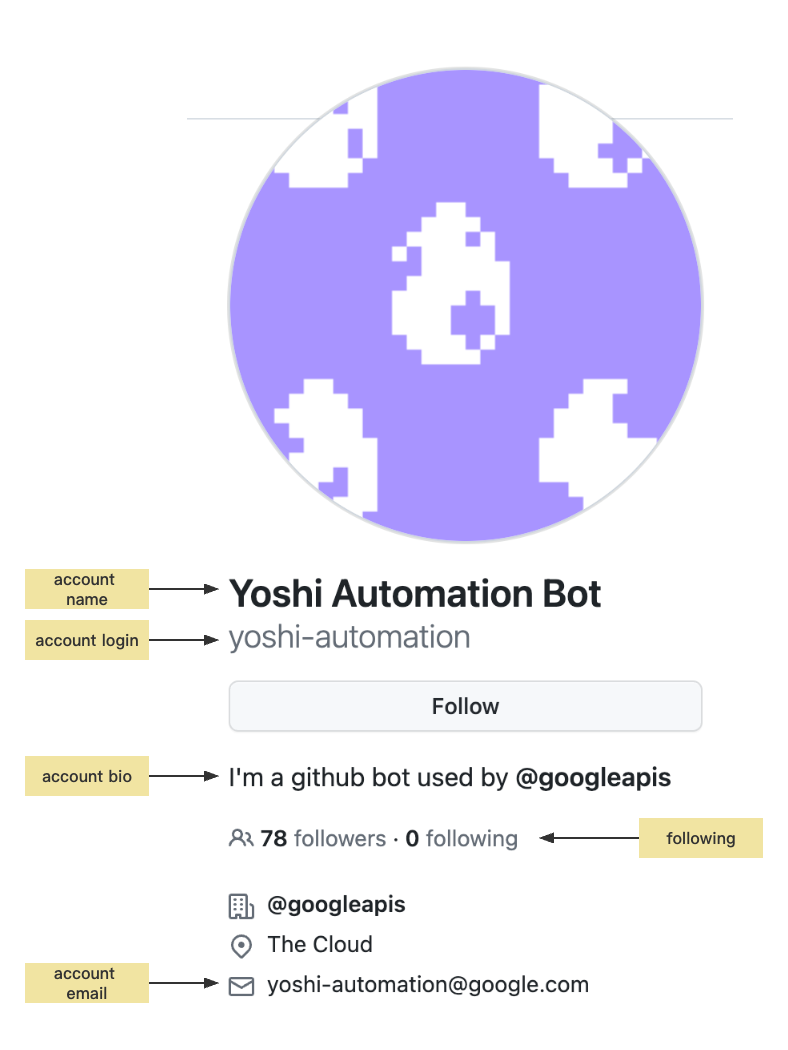}
\caption{Example of GitHub profile}
\label{fig:Example of GitHub profile.png}
\end{minipage}
\end{figure}

\subsection{Bot Behavior Analystic}

\begin{table}[htbp]
\centering
\begin{tabular}{|p{2.5cm}|p{4cm}|p{3cm}|p{5cm}|}
\hline
\textbf{Category} & \textbf{Description} & \textbf{Representative bot} & \textbf{Behavior} \\ \hline
\renewcommand{\arraystretch}{}
\multirow{4}*{\shortstack{Automatic\\Commenting\\Bot}} &  \multirow{4}{4cm}{Activate a comment on an issue followed by a textual response in the pull request comment once the user creates an issue, the pull request is accepted, or the CI/CD process is finalized.} & Repository Commander & Comment immediately under a newly created issue. \\ \cline{3-4} 
 &  & XRPL Bot & Comment immediately after being mentioned with "@." \\ \cline{3-4} 
 &  & quine-bot & Comment under the pull request after the user submits it. \\ \cline{3-4} 
 &  & Performance Testing Bot & When mentioned with "@" in the comments of a pull request, a comment will be published. \\ \hline
\multirow{4}{2.5cm}{Continuous Integration and Continuous Deployment/Delivery (CICD) Bot} & \multirow{4}{4cm}{Execute actions as part of the DevOps process post-PR submission to help facilitate workflow smoothness.} & GitHub Bot APP & Check if the information in the pull request meets the format requirements after it is submitted. \\ \cline{3-4} 
 &  & Mabl Bot & Display testing results in the checks section of the pull request. \\ \cline{3-4} 
 &  & Persona Features Bot & After a user's pull request is merged, a bot will submit a pull request to modify the CSV file. \\ \cline{3-4} 
 &  & Decca-Maven & Comment after a user submits a pull request to modify the dependency management script (i.e., pom.xml) or source code. \\ \hline
\multirow{4}{2.5cm}{Workflow Bot} & \multirow{4}{4cm}{Bots oversee the lifecycle of issues, pull requests, and discussions, which includes functions such as opening, closing, assigning, and labeling issues and pull requests.} & Boring Cyborg & Label pull requests by analyzing files modified in each PR. \\ \cline{3-4} 
 &  & Announcement Drafter & Creates a discussion based on information in the merged PR. \\ \cline{3-4} 
 &  & Paul the Alien & Streamlines GitHub work provides quick instructions like responding to comments, labeling, and merging PRs. \\ \cline{3-4} 
 &  & 0pdd.com & When a new PR is merged, an issue is generated if "@todo" appears anywhere in its comments. Once the code is resubmitted and the "@todo" is resolved, the corresponding issue is automatically deleted. \\ \hline
\multirow{4}{2.5cm}{Scanning Bot} & \multirow{4}{4cm}{Periodically or trigger-triggered scan the project's code files or related data, analyze their content.} & watchman-pypi & Trigger scans projects to create an issue. \\ \cline{3-4} 
 &  & open-digger bot & Reports weekly issue and star count statistics at a specific time every Monday. \\ \hline
\end{tabular}  
\caption{GitHub Apps behavior Category} 
\label{tab:GitHub Apps behavior Category}
\end{table}

Before diving into the feature selection process, it is essential to understand the characteristics of bot behavior within open-source software projects. By thoroughly analyzing bot activities, we can identify patterns and behaviors that differentiate bots from genuine users. In this subsection, we will explore various aspects of bot behavior, such as account creation, contribution patterns, and interaction with other users, which will inform our selection of meaningful features for bot detection.  

GitHub accounts can have an account tag designated as "Bot," which has limited permissions, specific repositories, and project access, and is limited to performing only specified tasks. Bot users can be authenticated through GitHub Apps or OAuth applications. They can interact with other GitHub users, such as creating code branches and leaving comments on issues or pull requests, among other actions. In our analysis, we focused on GitHub Apps in the GitHub Marketplace. GitHub Apps are applications that utilize the GitHub API to interact with code repositories. A GitHub account must be assigned to an App to operate and the assigned account type on GitHub will be recognized as a "Bot" type \cite{github-app-web}. 

Our analysis involved observing the behavioral patterns of 721 GitHub Apps. The GitHub Apps were classified into four different categories based on their behavioral characteristics as outlined in Table~\ref{tab:GitHub Apps behavior Category}:  (1) Automatic Commenting Bot: Certain bots can submit a comment by a textual response once the user creates an issue, the pull request is accepted, or the CI/CD process is finalized. For instance, the Repository Commander app will instantly comment on a welcome message for newly created issues. Additionally, XRPL Bot displays a comprehensive XRPL transaction description upon being @mentioned, while DepChecker Bot comments on any changes in npm dependencies after users submit their pull requests. These bots exhibit a tendency to generate a collection of nearly identical comments (i.e., they conform to predetermined comment patterns). Conversely, we discovered that comments made by human users are predominantly unique, with only a handful conforming to any existing pattern (such as “Thank you!”, “LGTM,” or “+1”). (2) Continuous Integration and Continuous Deployment/Delivery(CICD) Bot: Certain bots execute actions as part of the DevOps process post-PR submission to help facilitate workflow smoothness. For instance, the GitHub Bot APP retrieves metadata associated with pull requests, commits, branches, and trees, passing this information along to a user script for evaluation to assess the PR's viability for merging. Submissions that fail to adhere to requirements will receive a "Failed" status for their next submission attempt. The Mabl Bot APP embeds an end-to-end testing suite within GitHub's development workflow, inciting automatic testing and displaying test results after a PR is submitted. The Persona Features Bot listens for PR mergers, documents information, and subsequently submits an updated CSV file PR once it identifies a user's PR has been merged. (3) Workflow Bot: Bots oversee the lifecycle of issues, pull requests, and discussions, which includes functions such as opening, closing, assigning, and labeling issues and pull requests. Bots like Boring Cyborg app label pull requests by analyzing files modified in each PR. The announcement Drafter app creates a discussion based on information in the merged PR. Paul the Alien app streamlines GitHub work by providing quick instructions like responding to comments, labeling, and merging PRs.(4)Scanning Bot: These bots periodically or trigger-triggeredly scan the project's code files or related data, analyze their content, and generate related reports published on issues and pull requests. For example, watchman-pypi monitors dependency conflicts in millions of Python libraries in the PyPI ecosystem and trigger-scans projects to alert when there are dependency conflicts and to provide solutions. open-digger bot reports weekly issues and star count statistics at a specific time every Monday. While there are also some applications that do not fit into the category of bots. Some applications differ from bots because their activities occur outside of GitHub or do not involve any automated behavior. For instance, Ally enhances development visibility by retrieving crucial information from GitHub and displaying it in Slack, Microsoft Teams, Jira, Confluence, or dashboards, without automatization. The Slack + GitHub app seamlessly integrates code collaboration into conversations within Slack. This integration enables users to view and provide feedback on GitHub issues, pull requests, and code snippets without navigating outside of Slack. As a result, project management workflows are streamlined and team collaboration is improved. The app's functionality is limited to messaging and communications, and does not involve any periodic or automated activity.

Our findings unveiled some noteworthy patterns concerning user profile information from GitHub Apps-associated accounts. Specifically, our observations are as follows: (1) the bot accounts do not provide any details regarding their name, email, location, company, bio, and more. (2) The bots' name, bio, and email descriptions include terms such as bot, auto, ci, cla, code, io, logic, and assist. (3) Bots have few followers and followings, with most of them displaying a zero count. (4) The profile pictures are default images. Behavioral patterns and account  patterns set the foundation of our labeling approach. We used these patterns to handpick and meticulously analyze each label's attributes.  

\subsection{BotHawk dataset feature}

\begin{table}[h]
\centering
\begin{tabular}{|p{4cm}|p{4cm}|p{6cm}|}
\hline
\textbf{Dimensions} & \textbf{Features} & \textbf{Definition} \\ \hline
Profile Information & Account login & The primary identification of an account. \\ \cline{2-3}
 & Account name & The name of an account on GitHub. \\ \cline{2-3}
 & Account bio & The short bio description of an account. \\ \cline{2-3}
 & Account email & The email of an account. \\ \cline{2-3}
 & Account tag & Used to tag GitHub applications as "bot." \\ \cline{2-3}
 & Number of followings & The total number of users an account follows. \\ \cline{2-3}
 & Number of followers & The total number of users following the account. \\ \hline
Account Activities & Number of activity & Number of all activities an account has performed. \\ \cline{2-3}
 & Number of issues & Number of active issues of an account. \\ \cline{2-3}
 & Number of pull requests & Number of active pull requests of an account. \\ \cline{2-3}
 & Number of repositories & Number of active repositories of an account. \\ \cline{2-3}
 & Number of commits & Number of active commits of an account. \\ \cline{2-3}
 & Number of active days & Number of days the account was active in a year. \\ \cline{2-3}
 & Median response time & Median response time to the earliest event in issue or pull request. \\ \hline
Network features & Number of connection accounts & Number of accounts who have contact with this account. \\ \hline
Text Similarity & Comments similarity & Comments similarity created by an account. \\ \hline
Time Series & Periodicity of Activities & The trend of regular interval repetition of the account's activity over time. \\ \hline
\end{tabular}
\caption{An overview of features used to identify account type}
\label{table:An overview of features used to identify account type}
\end{table}

Based on our understanding of bot behavior and activities in open-source projects, we present the features extracted from the BotHawk dataset. These features encapsulate various dimensions of user behavior, enabling our model to effectively discern bots from genuine users. We selected 17 features grouped into five dimensions: profile information, account activities, text similarity, network features, and time series. Table~\ref{table:An overview of features used to identify account type} provides an overview of the dimensions and their associated features. In this subsection, we will describe each feature in detail and discuss its relevance to the bot detection task, providing a solid foundation for the following feature selection process.  

\begin{figure}[h]
\centering\includegraphics[width=1\linewidth]{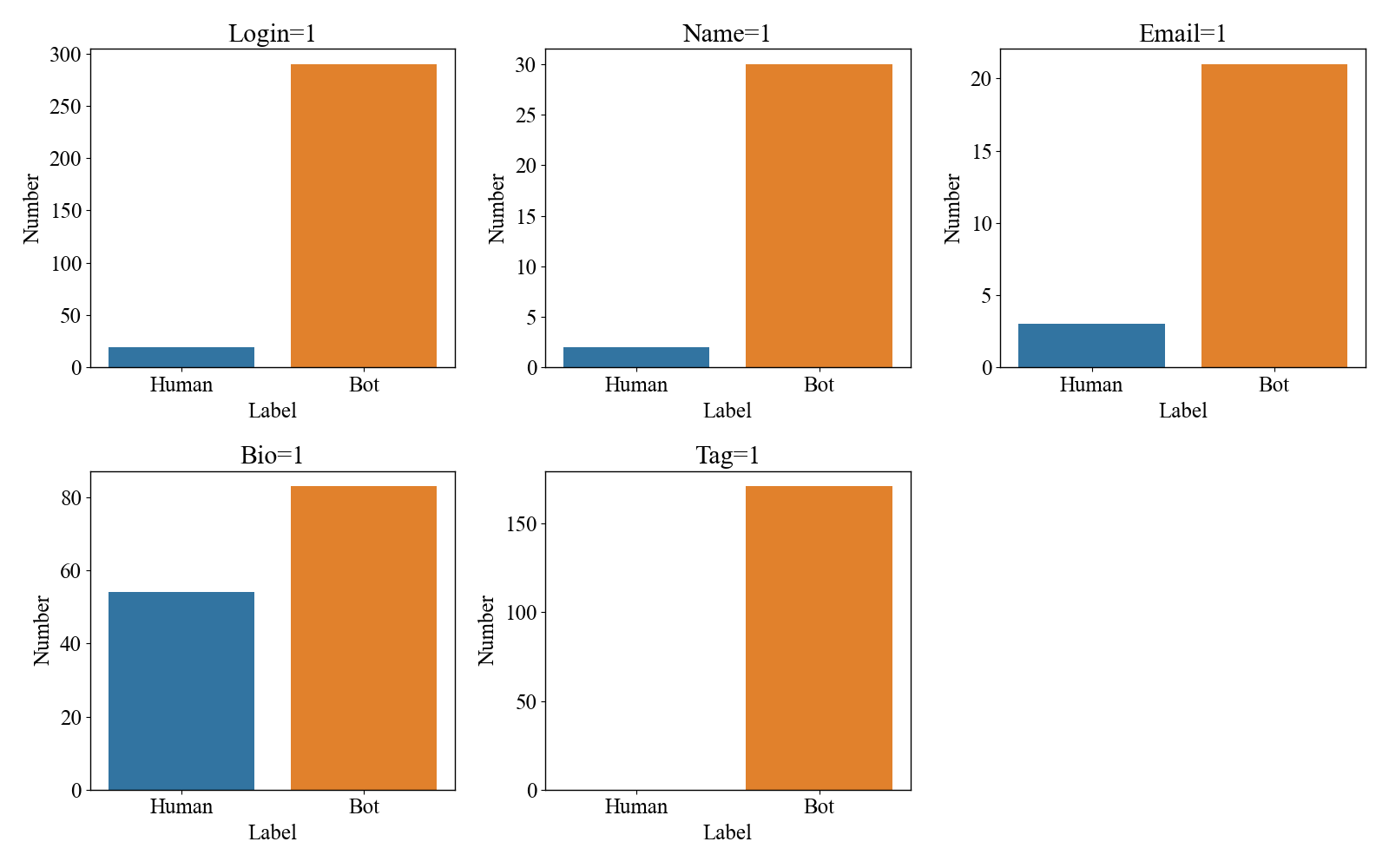}
\caption{Statistics of account information number between bot and human}
\label{fig:Statistics of account info number between bot and human}
\end{figure}

\textbf{Account login, Account name, Account bio, Account email:} In Figure~\ref{fig:Example of GitHub profile.png}, the personal introduction page for a GitHub account features the login name 'Yoshi Automation,' and users can supplement their profile with a profile name, as exemplified by 'Yoshi Automation Bot' in the same figure. The bio section on the introduction page is optional, allowing users to provide brief biographical summaries of themselves and details of their skills or interests. The email field also appears on the introduction page and displays an email address that a user has provided for contact. Through our observation, we have noted that bot profile accounts typically have certain substrings, including 'bot,' 'auto,' 'ci,' 'cla,' and 'io,' among others, in their login names, profile names, bios, and email addresses. Consequently, we considered the presence of these substrings in the text as a binary feature, with 0 for absence and 1 for presence, as shown in equation~\ref{eq:eq1}, which we encoded using label encoding. Figure~\ref{fig:Statistics of account info number between bot and human} provides evidence that The number of bot accounts containing the strings 'bot,' 'auto,' 'ci,' 'cla,' and 'io' in the login, name, and bio is significantly greater than the number of accounts labeled as human. This suggests that these strings are likely used to refer to accounts.  

\begin{equation}
\label{eq:eq1}
Feature_{login,name,bio,email} = 
\begin{cases}
1,  & \text{if account contains 'bot', 'auto', 'ci', 'cla', 'io', et.} \\
0, & \text{otherwise}
\end{cases}
\end{equation}

\textbf{Account tag:} GitHub has different types of accounts, each serving its distinct purposes. One such account type is the Bot tag, which is utilized for automating tasks such as code management, automated deployment, and automated testing. We have utilized label encoding to identify and differentiate this feature. This feature is labeled as 1 to indicate that an account has a Bot tag, as shown in equation~\ref{eq:eq2}, while 0 denotes an account that does not have a Bot tag, as observed in Figure~\ref{fig:Statistics of account info number between bot and human},  making Account tag a valuable feature for distinguishing between human and accounts.

\begin{equation}
\label{eq:eq2}
Feature_{tag} = 
\begin{cases}
1,  & \text{if account is 'Bot' tag} \\
0, & \text{otherwise}
\end{cases}
\end{equation}

\begin{figure}[h]
\centering\includegraphics[width=0.8\linewidth]{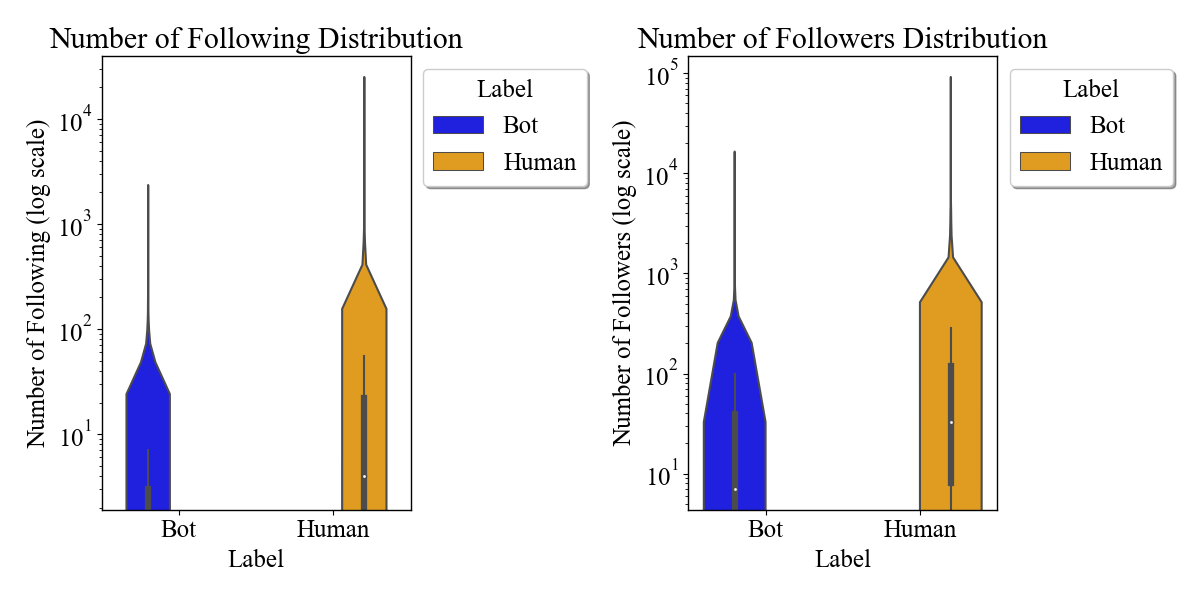}
\caption{Number of Following and Number of Followers Distribution}
\label{fig:Number of Following and Number of Followers Distribution}
\end{figure}

\textbf{Number of following, Number of follower:} Users typically follow other accounts on GitHub to stay informed of their activity \cite{blincoe2016understanding}. We contend that social activities such as these are more relevant to human users than bots, as humans are more likely to follow other accounts to keep abreast of recent activity, while accounts infrequently follow others. Therefore, we've utilized the number of followings and the number of followers as a feature to determine if accounts are likely operated by a human. Figure~\ref{fig:Number of Following and Number of Followers Distribution} illustrates the distribution of following and followers, and the logarithmic form is utilized to represent the data as some accounts have a vast number of followers and following. The data indicate that human accounts tend to have significantly higher followers and following than Bot accounts.

\begin{figure}[h]
\centering\includegraphics[width=1\linewidth]{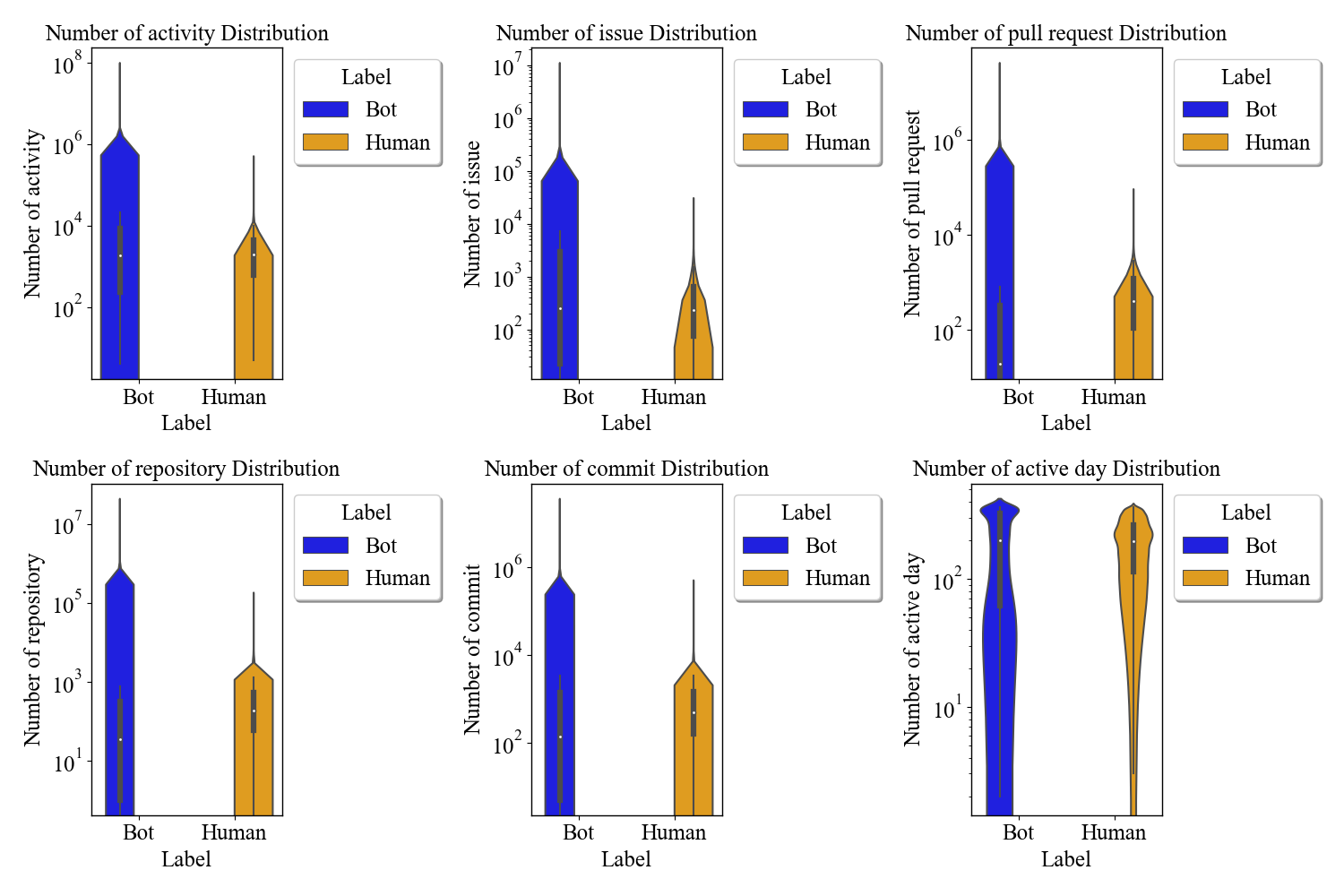}
\caption{Acticity Issue PR Repository Commit Activity per day Distribution}
\label{fig:Acticity Issue PR Repository Commit Activity per day Distribution}
\end{figure}

\textbf{Number of activities, Number of issues, Number of pull requests, Number of repositories, Number of commits, Number of active days:} Activity encompasses the actions of users on code hosting platforms, including creating, merging, or closing pull requests, writing comments, and submitting code. These platforms also allow users to report and discuss issues in software projects, which helps identify and solve problems like bugs, functional requirements, and suggested improvements. On the other hand, a pull request is a process where code changes are submitted to be merged into another code repository. Meanwhile, a repository is where code, documents, images, and other files are stored on the platform, also referred to as a source code library. Whenever changes are made to the code repository, the process is called committing, and these changes can also be treated as version or history records of the repository. Bot activities are fixed, repetitive, and predetermined. In our manual analysis, we discovered that bot accounts perform more daily activities than humans, which we expected given their role in automated repetitive tasks that reduce developers' costs. To distinguish between bot and human accounts, we use features such as overall activity counts, activity counts specifically tied to repositories, issues, pull requests, commits, and active days. As shown in Figure~\ref{fig:Acticity Issue PR Repository Commit Activity per day Distribution}, bot accounts tend to have considerably higher active event counts than those of human accounts, and the numbers of issues, pull requests (PRs), repositories, and commits also have similar patterns. In contrast, human accounts show a more uniform distribution. These findings are consistent with previous research on social coding and bot behavior in GitHub \cite{dabbish2012social}. Moreover, our results can contribute to improving automated testing and fault prediction in open-source software development \cite{gyimothy2005empirical}.

\textbf{Median response time:} Our analysis also shows that bot accounts may respond more promptly than human accounts. This is due to the fact that bots operate as monitoring tools in code repositories and are activated based on predefined actions (such as creating issues or pulling requests). We consider the earliest event on the issue or pull request timeline an essential indicator of the type of account response. We calculated the median response time for an account that responded to the same issue or pull request whether or not the account was mentioned, as in equation~\ref{eq:fea_mrt}, in which $t_n$ is the responding time and $t_{n-1}$ is last event time to the same issue or pull request. The statistical analysis of Median Response Time Distribution, exemplified in Figure~\ref{fig:Number of connection accounts, Median response time Distribution, Periodicity of Activities Distribution}, depicts the logarithm of the response time. Our findings specify that most Bot accounts have a quicker response time.

\begin{figure}[h]
\centering\includegraphics[width=1\linewidth]{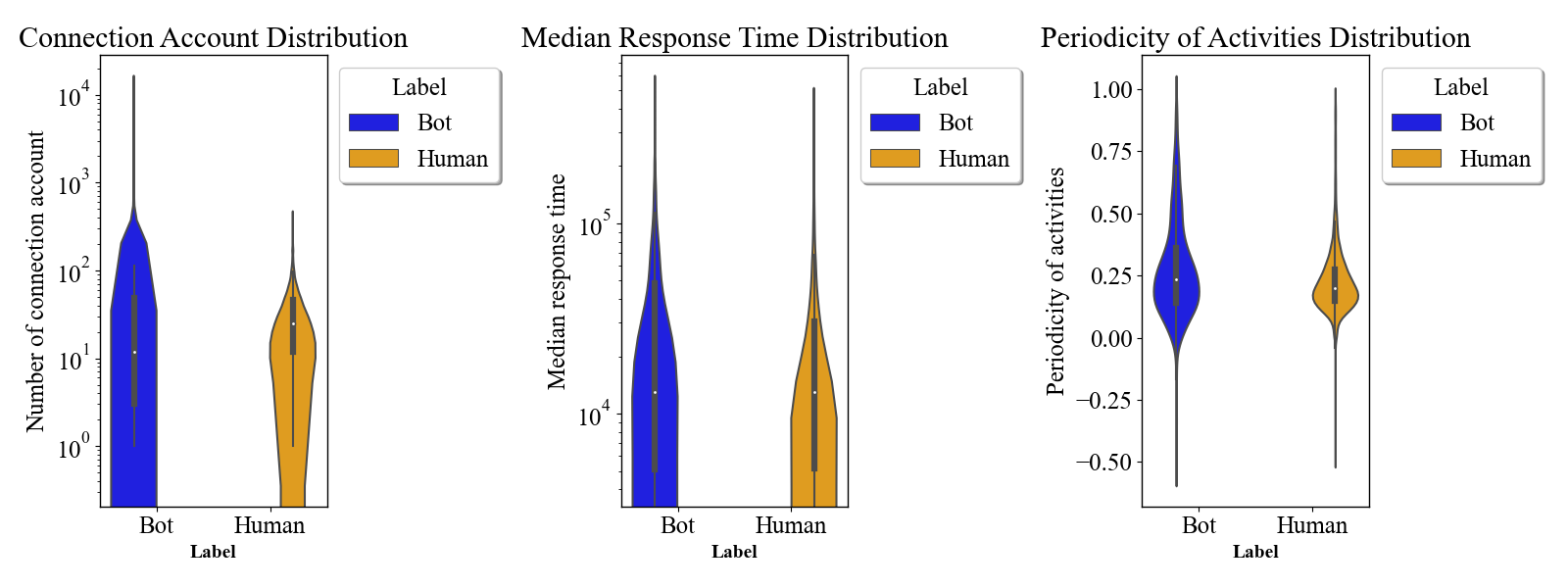}
\caption{Number of connection account, Median response time Distribution, Periodicity of Activities Distribution}
\label{fig:Number of connection accounts, Median response time Distribution, Periodicity of Activities Distribution}
\end{figure}

\begin{equation}
\label{eq:fea_mrt}
Feature_{Median\ Response\ Time} = median({t_n - t_{n-1}})
\end{equation}

\begin{figure}[h]
\centering\includegraphics[width=0.6\linewidth]{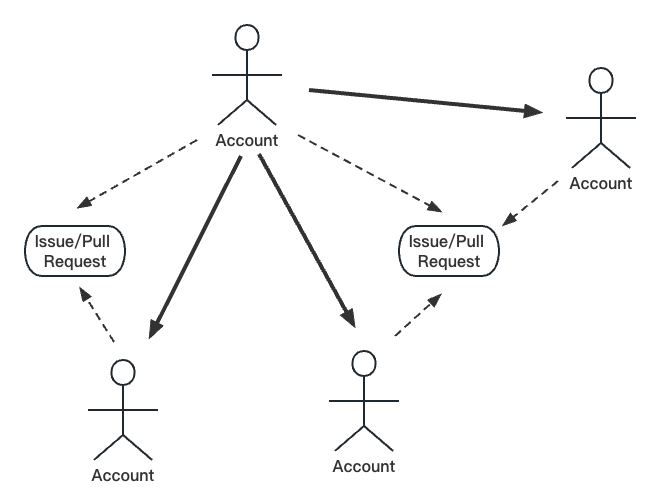}
\caption{Network of interconnected accounts}
\label{fig:Network of interconnected accounts}
\end{figure}

\textbf{Number of connection account:} Bot accounts take part in several activity events (Issue, PR), through which they establish social connections with multiple accounts. Hence, we make use of the number of accounts that have created social bonds with the account via issues and PRs. Figure~\ref{fig:Network of interconnected accounts} illustrates a network of interconnected accounts. Dashed lines indicate the activation of an account through opening or commenting on an issue or a pull request. Solid lines establish a connection between an account and other related accounts. Figure~\ref{fig:Number of connection account, Median response time Distribution, Periodicity of Activities Distribution} demonstrates that Bot accounts have more connected users, while Human accounts have numbers that are typically nearer to the median. We attained these findings by taking the logarithm of the number of connected users.

\textbf{Text Similarity:} Our analysis of behavioral patterns revealed that bots tend to generate comments after specific operations. These comments contain analysis reports (including test code coverage), welcome messages, and operation content, among others. Most of these comments use predefined templates, leading to a high degree of similarity. We consider the similarity of comment text as a critical feature. There are three common text similarity algorithms: word frequency statistics, vector space models, and semantic analysis. After analyzing comments made by bots, we observed that they primarily use templates. Therefore, semantic similarity is not a suitable approach for our research. Therefore, we selected Jaccard similarity algorithm, cosine similarity algorithm, and TF-IDF similarity algorithm. We computed the similarity scores between each user's most recent 100 comments and then calculated the average. Our results, presented in Figure~\ref{fig:Jaccard, TF-IDF, Cosin Smilarity Distribution}, indicate that although the disparity in performance is minimal, the TF-IDF similarity algorithm performs slightly better. Thus, we adopt the TF-IDF similarity algorithm for our research.

\begin{figure}[h]
\centering\includegraphics[width=\linewidth]{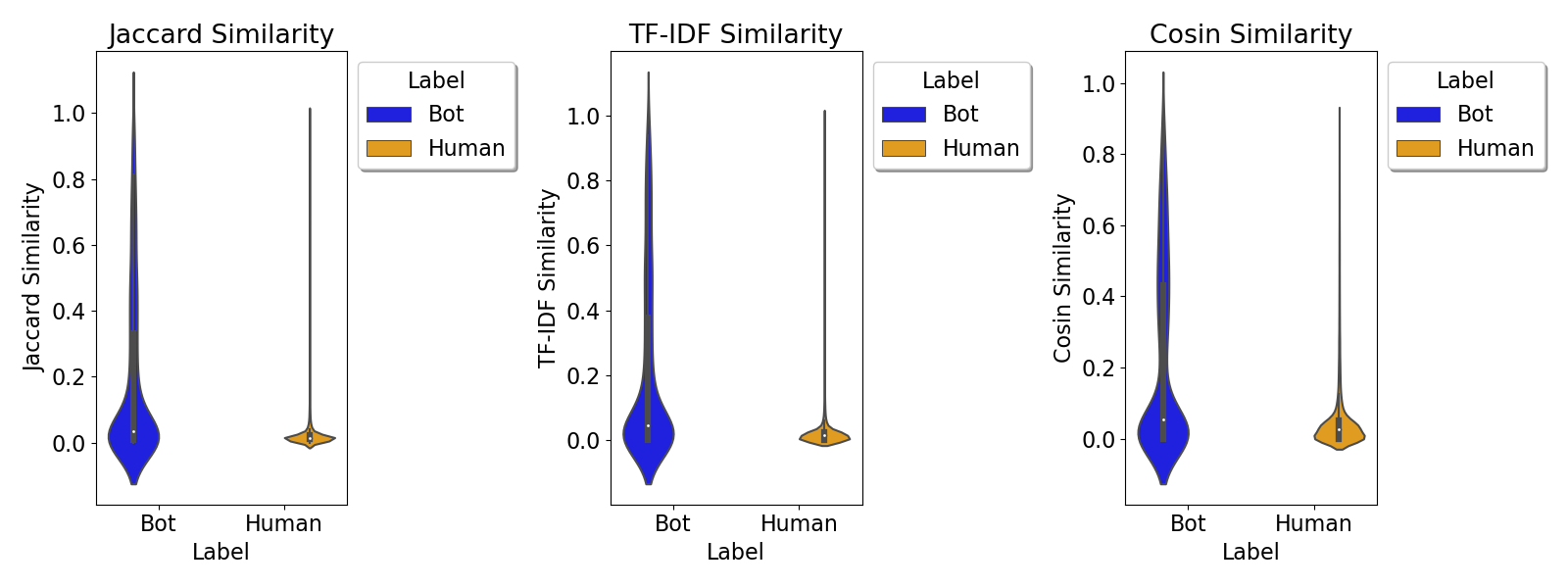}
\caption{Jaccard, TF-IDF, Cosin Smilarity Distribution}
\label{fig:Jaccard, TF-IDF, Cosin Smilarity Distribution}
\end{figure}

Through our rigorous feature selection process, we have identified a set of highly relevant and informative features that allow our BotHawk model to effectively discriminate between bots and genuine users, enhancing its accuracy and interpretability.

\section{Classification model}
\label{S:5}

In this section, we present the BotHawk classification model, which aims to detect bots in open-source software projects. We employ an ensemble learning approach to train our model using the features selected in Section 4 and evaluate the performance based on the ground truth dataset described in Section 3.

\subsection{Introduction}

We address the bot detection problem as a binary classification task, where the goal is to distinguish between two classes: bots and genuine users. In this context, we represent each account in our dataset as a feature vector $ X $, with elements $x_i (i=1,2,...,17)$, corresponding to the 17 features extracted from the accounts' behavior and interactions in open-source software projects. The target variable Y represents the account type, with Y=1 denoting a bot account and Y=0 denoting a genuine user account. Formally, given a dataset $ D = {(X_1, Y_1), (X_2, Y_2), ..., (X_n, Y_n)} $, where $ X_i \in  R^{17} $ and $ Y_i \in  {0,1} $, our objective is to learn a classification model $ f(X) $ that can accurately predict the target variable Y, i.e., $f(X_i) \approx Y_i$ for all i in the dataset. The classification model $ f(X) $ assigns a probability $ p(Y=1|X) $ to each account, representing the likelihood that the account is a bot.  

A common approach to binary classification is to use a decision function g(X) that maps the feature space to a real number and then applies a threshold T to determine the class label:

\begin{equation}
\label{eq:pred}
Y_{\text{pred}} = 
\begin{cases}
1 & \text{if } g(X) \geq T \\
0 & \text{otherwise}
\end{cases}
\end{equation}

In bot detection, g(X) can be interpreted as a measure of the bot possibility for each account. By comparing g(X) to the threshold T, we can classify each account as either a bot $ (Y_{pred} = 1) $ or a genuine user $ (Y_{pred} = 0) $. We aim to learn a classification model f(X) that can accurately estimate the decision function g(X) and minimize the classification error on unseen data.  

We have explored a range of base models, including Logistic Regression, Random Forest, Naive Bayes, etc. An overview of three classification algorithms is provided, along with an explanation of the reasoning underpinning their selection for our comparative analysis. Logistic Regression is a well-behaved classifier for roughly linear features or linearly separable data \cite{peng2002introduction}. It can handle nonlinear data through feature discretization and mapping, is robust to noise, and can avoid overfitting with L1 or L2 regularization techniques. On the other hand, tree ensembles like Random Forests are a collection of decision trees and do not require linear features, making them ideal for handling certain data types \cite{kocev2013tree}.  Naive Bayes is a probabilistic classifier based on Bayes' theorem that assumes feature independence \cite{rish2001empirical}. This assumption allows for faster convergence and fewer data in training compared to discriminative models like Logistic Regression but can be impractical in real-world scenarios. Nonetheless, Naive Bayes often performs well in practice.  

\subsection{Methodology}

We present a comprehensive approach to constructing an efficient and accurate ensemble model for bot detection in open-source software projects. Our methodology encompasses a series of steps, starting from data preprocessing to model evaluation, to ensure the best possible performance while maintaining model simplicity.

Our strategy involves training various base models, including support vector machine, decision tree, random forest, K-nearest neighbors, logistic regression, Naive Bayes, and neural networks, on a preprocessed dataset that has undergone encoding and normalization. Table~\ref{table:Base Model Evaluation Metrics} shows that the F1-score and AUC values of the base classification algorithms are quite low owing to the dataset imbalance. Hence, we have utilized undersampling and ensemble learning methods to optimize the model for identifying bot accounts.

\begin{table}[h]
\centering
\caption{Base Model Evaluation Metrics}
\begin{tabular}{|l|c|c|c|c|c|}
\hline
\textbf{Model} & \textbf{Accuracy} & \textbf{Precision} & \textbf{Recall} & \textbf{F1-Score} & \textbf{ROC-AUC} \\ \hline
Logistic Regression & 0.909 & 0.385 & 0.590 & 0.466 & 0.574 \\
Decision Tree Classifier & 0.791 & 0.213 & 0.782 & 0.335 & 0.505 \\
Support Vector Classifier & 0.883 & 0.323 & 0.677 & 0.437 & 0.536 \\
Gaussian Naive Bayes & 0.952 & 0.698 & 0.496 & 0.580 & 0.526 \\
K Nearest Neighbors & 0.823 & 0.226 & 0.677 & 0.339 & 0.517 \\
Random Forest Classifier & 0.879 & 0.340 & 0.846 & 0.485 & 0.639 \\ \hline
\end{tabular}
\label{table:Base Model Evaluation Metrics}
\end{table}

Bagging is a typical ensemble learning algorithm proposed by Breiman \cite{breiman1996bagging}.  In brief, Bagging generates a training set $X^t=(X^t_1,X^t_2,...,X^t_n)$ of the same size as the original sample set $X=(X_1,X_2,...,X_n)$ by using bootstrap sampling method \cite{johnson2001introduction}, , which samples the original sample set $X=(X_1,X_2,...,X_n)$ with replacement uniformly. By training through multiple sampling, a series of weak classifiers $f_1,f_2,...,f_m$ are obtained, and the final strong classifier is obtained by simple averaging.

\begin{figure}[h]
\centering\includegraphics[width=\linewidth]{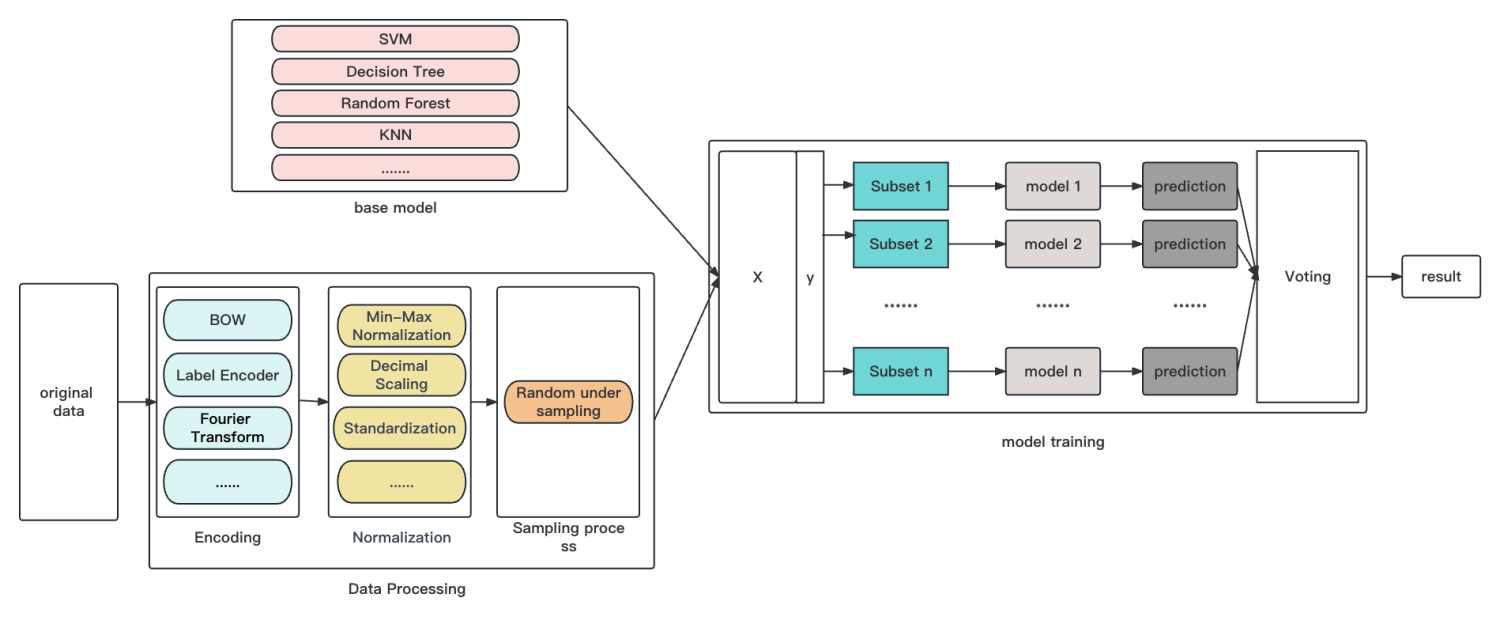}
\caption{Workflow for BotHawk Model}
\label{fig:Workflow for BotHawk Model}
\end{figure}

Our BotHawk bot detection model's training process involves both data processing and model training, As shown in Figure~\ref{fig:Workflow for BotHawk Model}. We addressed the data imbalance issue by using a compact and efficient ensemble model that combines different weak classifiers. As a result, we achieve superior bot detection performance in open-source software projects.

The dataset features are heterogeneous, comprised of numerical and categorical values. The numerical features include the number of activities, periodicity, and others, while the categorical features indicate whether a feature is related to bots. The numerical features were normalized or standardized to reduce the impact of differences in range and units on model performance. Categorical features were encoded to ensure proper handling by the model. However, a single classifier struggles to utilize information from multiple sources and cannot achieve satisfactory classification performance for bots, regardless of the amount of available data. 

By integrating the encoder and normalization steps, we ensure that the input features are adequately prepared during the modeling process, enabling both the base and ensemble models to learn from the data and achieve high predictive performance effectively. To create the ensemble model, our base model encompasses several widely used classification algorithms, including Support Vector Machines (SVM), Decision Trees, Random Forests, K-Nearest Neighbors, Logistic Regression, and Naive Bayes. We employ a voting algorithm to leverage the strengths of the individual weak models. The voting algorithm operates by aggregating the individual predictions made by the base models for a given input data point. We utilize a hard voting strategy where each base model independently predicts and produces its own result, and the final prediction is determined by majority voting. Therefore, in the Bagging algorithm, the base models are independent of each other, with equal weights for each model, and the final result is determined by the majority vote of the predicted results from each base model. Given the importance of result reliability and stability, we performed 5 iterations for each algorithm. In each iteration, we used GridSearch and cross-validation to evaluate the performance of each parameter combination, effectively measuring the impact of different parameter combinations on model performance. After completing the 5 iterations, we analyzed the optimal performance achievable by each algorithm.  

The results of this analysis can be found in Table~\ref{tab:bagging_xgboost_eval}. The table shows the performance of various machine learning models, including bagging algorithms applied to five different models and the XGBoost model. Bagging-RandomForest achieves the best performance with an accuracy of 0.880, precision of 0.893, recall of 0.871, F1-score of 0.882, and an AUC value of 0.947. The XGBoost model also performs well, with an accuracy of 0.870 and AUC of 0.944. Based on these results, we conclude that bagging-RandomForest is the best-performing model with the most potential for practical use in this specific dataset and problem domain.  

\begin{table}[h]
\centering
\caption{Performance of bagging model with differenf base models}
\begin{tabular}{|l|c|c|c|c|c|}
\hline
\textbf{Model} & \textbf{Accuracy} & \textbf{Precision} & \textbf{Recall} & \textbf{F1-score} & \textbf{AUC} \\
\hline
Bagging-DecisionTree & 0.859 & 0.876 & 0.847 & 0.861 & 0.930 \\
Bagging-KNeighbors & 0.814 & 0.880 & 0.739 & 0.803 & 0.891 \\
Bagging-RandomForest & 0.880 & 0.893 & 0.871 & 0.882 & 0.947 \\
Bagging-LogisticRegression & 0.747 & 0.826 & 0.647 & 0.725 & 0.891 \\
Bagging-SVM & 0.810 & 0.883 & 0.727 & 0.797 & 0.884 \\
Bagging-GaussianNB & 0.745 & 0.977 & 0.518 & 0.677 & 0.865 \\
XGBoost & 0.870 & 0.891 & 0.851 & 0.871 & 0.944 \\
\hline
\end{tabular}
\label{tab:bagging_xgboost_eval}
\end{table}

By combining the random forest models through the voting algorithm, we aim to construct a powerful ensemble model that capitalizes on the strengths of each individual model, providing improved bot detection performance in open-source software projects.

\subsection{Result}

\begin{figure}[h]
\centering\includegraphics[width=1\linewidth]{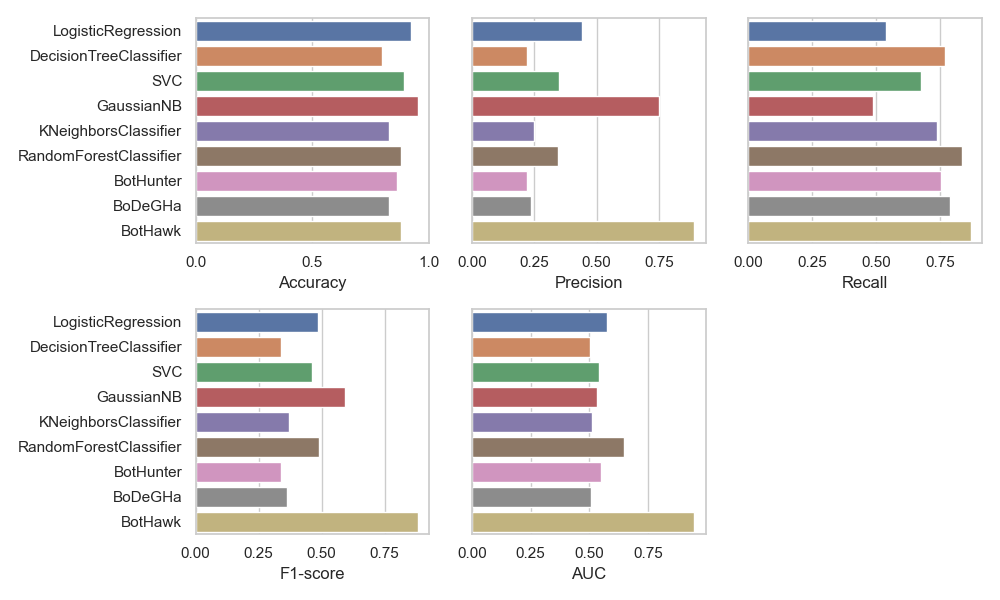}
\caption{Comparison of Classification Models}
\label{fig:Comparison of Classification Models}
\end{figure}

\begin{figure}[h]
\centering\includegraphics[width=0.6\linewidth]{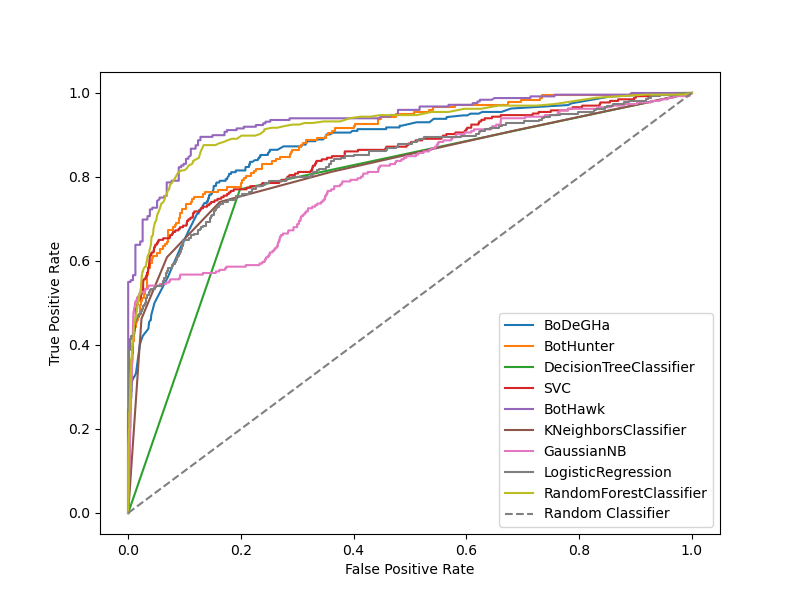}
\caption{ROC Curve for Different Models}
\label{fig:ROC Curve for Different Models}
\end{figure}

In the RESULTS section of our paper, we begin by discussing the evaluation metrics we used to assess the performance of our ensemble model for bot detection in open-source software projects. The chosen evaluation metrics include accuracy, precision, recall, and F1-score. Due to the highly imbalanced dataset, we used AUC as the primary performance metric. AUC provides an aggregated performance measure across all classification thresholds and represents the probability that randomly selecting a positive sample will have a higher rank than randomly choosing a negative sample. Additionally, we listed accuracy, precision, recall, and F1-score as our evaluation metrics for classification. For all experiments, we employed grid search and 5-fold cross-validation to measure the model's performance and recorded the average value of each performance metric.

The result is shown in Figure~\ref{fig:Comparison of Classification Models}. The first indicator is Accuracy, which represents the proportion of all correctly classified samples to the total number of samples. The top three most accurate models, in descending order, are BotHawk (0.98), BotHunter (0.97), and GaussianNB (0.95). BotHawk has the highest accuracy among the models, indicating that this model has higher accuracy and reliability in sample classification. The models with the highest precision were GaussianNB (0.728), BotHunter (0.881), and BoDeGHa (0.804). And the top three models for recall, in descending order, are BotHawk (0.806), DecisionTreeClassifier (0.776), and SVC (0.681). Additionally, the F1-Score indicator combines the precision and recall indicators. BotHawk has the best performance, with an F1-Score metric of 0.89, significantly surpassing all other models. Finally, in terms of the AUC metric, the three models with the highest performance are BotHawk (0.947), RandomForestClassifier (0.639), and SVC (0.536). The ROC curve for BotHawk is closer to the top-left corner than other models, highlighting its superior performance on the dataset and potential as a novel approach to detecting positive samples in this domain. as presented in Figure~\ref{fig:ROC Curve for Different Models}. The results suggest that BotHawk exhibits the highest AUC value, indicating its superior ability to distinguish between positive and negative samples. However, it is worth noting that the performance of these models can vary depending on the specific problem domain and dataset used, making it crucial to evaluate and compare the models carefully before selecting the most appropriate one for a given task.  

BotHawk performs the best overall, with a strong performance on the key indicators. BotHawk performs strongly on the AUC and F1-Score indicators with the value is 0.947 and 0.89.

\section{Discussion}
\label{S:6}
\textbf{Identifying bot account issues:} Ideally, BotHawk effectively differentiates between  bot accounts and human accounts. To achieve this, BotHawk is trained on a dataset encompassing a wider variety of bot accounts, offering a more realistic portrayal of bot-related situations. However, BoDeGHa performs well in identifying bot accounts that exhibit comment-related features, as it is primarily trained based on such characteristics. Nevertheless, BoDeGHa is limited in its scope to assessing bot behavior solely within a specific repository, lacking a more comprehensive perspective in identifying bots. On the other hand, BotHunter concentrates on simplistic features and fails to explore the comprehensive behavioral characteristics associated with bots. These factors contribute to BotHawk's exceptional performance when handling datasets that closely emulate real-world scenarios. We found that the recognition performance of CICD and Scanning bots is superior to that of other models. BotHawk identified the 'dotnet-bot' account as a bot account, whereas other models identified it as a human account. However, upon analyzing the account, many automated pull request actions were discovered belonging to the CICD bot. The reason for this is the adoption of network features and Time Series features, which can encompass a broader range of bot types.

\begin{figure}[htbp]
\centering
\begin{minipage}[t]{0.48\linewidth}
\centering
\includegraphics[width=\linewidth]{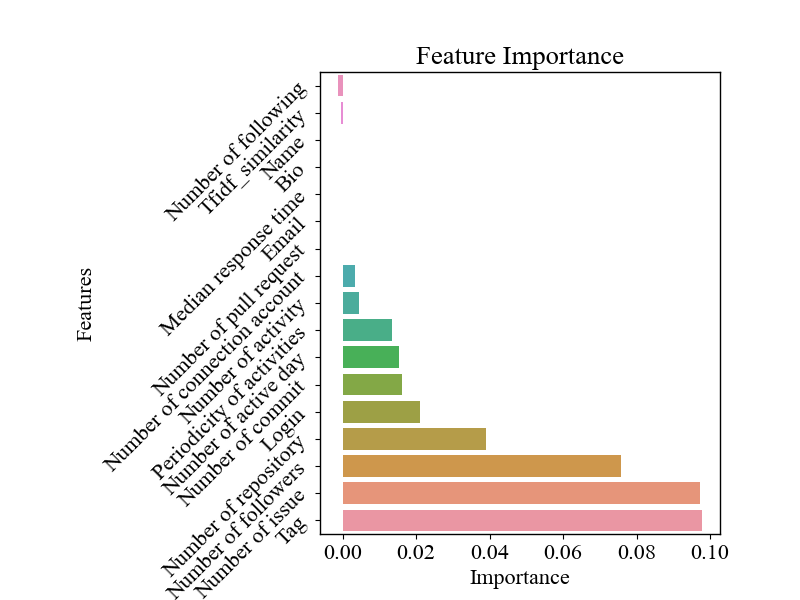}
\caption{Feature Importance for BotHawk model}
\label{fig:Feature Importance for BotHawk model}
\end{minipage}
\hfill
\begin{minipage}[t]{0.48\linewidth}
\centering
\includegraphics[width=\linewidth]{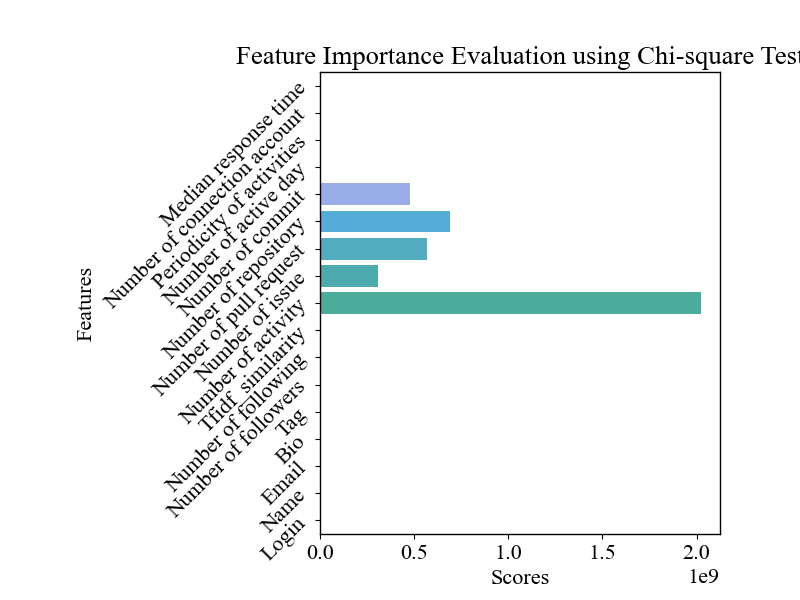}
\caption{Feature Importance Evaluation using Chi-square Test}
\label{fig:Feature Importance Evaluation using Chi-square Test}
\end{minipage}
\end{figure}

\textbf{Importance of features: }The feature importance plot of the BotHawk model is utilized to elucidate the significance of each feature. This tool serves as a visual aid for understanding the features that exert the greatest impact on the model's predictions and offers insights into interpreting these predictions. Figure~\ref{fig:Feature Importance for BotHawk model} illustrates that features such as 'tag' (0.0977), 'Number of followers' (0.0758), and 'Number of Issues' (0.0973) exhibit higher levels of importance. This implies a robust positive correlation between the identification performance of open-source bots.  

The 'tag' feature represents an official recognition of bot characteristics, while increased attention and involvement in issues may indicate the most discernible attributes of open-source bot accounts. Conversely, features like 'Number of Activities,' 'Text Similarity,' and 'Number of Connection Accounts' hold lesser significance, suggesting a diminished influence on the identification performance of bots within this specific dataset. This indicates that prior research placed excessive reliance on text similarity features, thus oversimplifying the problem without undertaking a comprehensive perspective.  

Additionally, bot accounts and human users on coding platforms can have a high level of activity and connections with different developers, which further emphasizes the increasing importance of collaborative behavior in open-source software. On the other hand, the 'following' feature (-0.0012)  and 'text similarity' feature(-0.0004) exhibit a negative impact or correlation with the predictions of open-source bot accounts. We believe this feature to be redundant, as it does not provide additional information during the model training process and may introduce noise, thereby decreasing the model's performance.  

Out of the list of features, the importance values for name, email, bio, Median Response Time, Number of Pull Requests, and Periodicity of Activities are all equal to 0. This suggests that these features have little to impact on the model's predictions, meaning they are not relevant or useful in predicting the outcomes in this context.  

The chi-squared test is a statistical method used to determine variables' correlation. Figure~\ref{fig:Feature Importance Evaluation using Chi-square Test} indicates that the number of followers, number of fans, periodicity of activities, and active days are potentially linked to our predicted target variable. There may be less correlation of the target variable with the number of repositories, number of pull requests, activity level, and number of issues. Other variables could have a minimal effect, and it is improbable for these variables to impact the target variable significantly.   

Upon scrutiny of the chi-squared test and feature importance analysis, we can infer that the number of followers, number of repositories, and tags variables may have the highest significance to our target variable, while the remaining variables have nominal importance. Nevertheless, we must consider that these conclusions are not definitive and could vary based on different datasets, modeling methods, and models.  

\section{CONCLUSION}

Detecting OSS bot accounts is a challenging task. It is essential to identify robot accounts in order to effectively develop and operate open source software. In this study, our primary focus was to detect bots in open-source software projects and provide valuable insights into their behavior. Our contributions are four-fold. Firstly, we created a standardized dataset named BotHawk dataset, which contains 19,779 accounts and 17 properties. We follow a systematic workflow to ensure data accuracy, generalization ability, scalability, and timeliness. This dataset will assist the research community in further studying bots in open-source projects. Secondly, we analyzed and classified open-source bots' behavioral patterns into four categories: Automatic Commenting Bot, Continuous Integration and Continuous Deployment/Delivery (CI/CD) Bot, Workflow Bot, and Scanning Bot. This classification may help researchers better understand bot types and their different roles in open-source software projects. Thirdly, we identified 17 features that can differentiate bot accounts from human accounts, providing valuable insights into bot behavior in open-source projects. Lastly, we proposed an ensemble learning model, surpassing traditional machine learning models and existing bot detection methods. We analyzed the comparison between the model and state-of-the-art models and the importance level of the features. The task of OSS bot detection is challenging due to dataset alignment with real-world scenarios. Our proposed model combines multiple base model strengths, and as a result, it provides a more precise and robust solution for detecting bot accounts in open-source software projects. Furthermore, we find that the number of followers, number of repositories, and tags contain the most relevant features to identify the account type. We plan to use our classification model in the social-technical empirical analysis of collaborative software development.

Our study contributes to understanding bot activities in open-source projects and provides practical methods for detecting bots. Our findings and proposed ensemble learning model may assist future research in exploring the impact of bots on open-source communities and developing effective strategies for managing their activities.  

Besides the contributions mentioned earlier, there are several future directions for bot detection in open-source software projects:  

\textbf{Bot Detection Tool:} One promising direction is to develop a practical bot detection tool that can easily integrate into open-source project platforms, such as GitHub, GitLab, or Bitbucket. This tool can assist project contributors and maintainers monitor bot activities, leading to better decision-making about bot usage and understanding their impact on project progress.  

\textbf{Incorporation of Graph Features:} Another promising direction for future research is to incorporate graph features into bot behavior analysis. By studying the relationships and interactions between users and bots and their connections within the open-source community, we can better understand bot roles in project ecosystems. Such an approach could potentially improve the performance of our bot detection model and provide new insights into bot behavior patterns.   

\textbf{Classification of Types of OSS bot Accounts:} Bot accounts can be categorized into various types based on different dimensions. Subsequently, employing multi-label classification techniques allows for a more refined classification of OSS bot account types. This study investigates the impact of various bot types on open-source repository projects. 

In summary, exploring new research areas related to bot detection in open-source software projects is essential. By pursuing these future directions, we can continue to advance our understanding of bot activities and develop more effective strategies for managing and leveraging their contributions to the open-source community.





\bibliographystyle{elsarticle-num}
\bibliography{sample}







\end{document}